\documentstyle[procl]{article}

\input{psfig.sty}

\def\Journal#1#2#3#4{(#1) {#2} {\bf #3}, #4}

\def\AAp{\em Astron.~Astrophys.}

\def\ApJ{\em Astrophys.~J.}
\def\ApJL{\em Astrophys.~J., Lett.}

\def\ARAaAp{\em Annnu.~Rev.~Astron.~Astrophys.}
\def\MNRAS{\em Mon.~Not.~R.~Astron.~Soc.}
\def\Nat{\em Nature\/}

 \newcommand{\half}{\textstyle{\frac12}}
\newcommand{\HI}{{\rm H\,\scriptstyle I}}

\newcommand{\la}{\;
  \raise0.3ex\hbox{$<$\kern-0.75em\raise-1.1ex\hbox{$\sim$
  }}\;\hskip-2pt }
\newcommand{\ga}{\;
  \raise0.3ex\hbox{$>$\kern-0.75em\raise-1.1ex\hbox{$\sim$
  }}\;\hskip-2pt }

\newcommand\vect[1]{ \mathchoice
    {\hbox{\boldmath$\displaystyle{#1}$\unboldmath}}%
    {\hbox{\boldmath$\textstyle{#1}$\unboldmath}}%
    {\hbox{\boldmath$\scriptstyle{#1}$\unboldmath}}%
    {\hbox{\boldmath$\scriptscriptstyle{#1}$\unboldmath}}}

\newcommand{\kms}{\,{\rm km\,s^{-1}}}

\newcommand{\cmcube}{\,{\rm cm^{-3}}}

\newcommand{\mkG}{\,\mu{\rm G}}

\newcommand{\K}{\,{\rm K}}
\newcommand{\kpc}{\,{\rm kpc}}
\newcommand{\ppc}{\,{\rm pc}}
\newcommand{\yr}{\,{\rm yr}}

\begin{document}


\title{A Dynamic Interstellar Medium:\\ Recent Numerical Simulations}

\author{Anvar Shukurov}

\address{Department of Mathematics, University of
Newcastle,\\ Newcastle upon Tyne NE1~7RU, England}

\maketitle

\abstract{Recent numerical simulations of the interstellar medium
driven by energy input from supernovae and stellar winds indicate
that $\HI$ clouds can be formed by compression in shock waves and
colliding turbulent streams without any help from thermal and
gravitational instabilities. The filling factor of the hot phase
in these models does not exceed 20--40\% at the midplane. The
hot gas is involved in a systematic vertical outflow at $|z|<1$--3\,kpc, similar to that
expected for galactic fountains, whereas the warm component may remain in hydrostatic
equilibrium. The turbulent velocity is larger in the warmer phases, being $3, 10$ and
$40\kms$ in the cool, warm and hot phases, respectively, according to one of the
simulations. The models exhibit global variability in the total kinetic and thermal
energy and star formation rate at periods of (0.4--$4)\times10^8\yr$. Current models are
still unable to reproduce dynamo action in the interstellar gas; we briefly discuss
implications of the dynamo theory for turbulent interstellar magnetic fields.}

\section{Introduction}
Recent numerical simulations of the interstellar medium (ISM) heated by supernovae and
stellar winds have highlighted the importance of numerical experiment in
the studies of the multiphase ISM. The simulations play a complimentary r\^ole to
observational and semi-analytical studies and are especially important in clarifying such
long-standing problems as the filling factors of the various phases of the ISM, the nature
and lifetimes of $\HI$ clouds, dynamo action, etc. Simulations capturing relevant
physical effects at appropriate resolution have become feasible only recently, and even
first results reviewed here provide insight into many controversial issues.

The range of gas temperatures and densities in the ISM is tantalizingly
wide, from $T\simeq10^6\K$, $n\simeq10^{-3}\cmcube$ in the hot phase to 10\,K and
more than $10^3\cmcube$ in molecular clouds. The gas is involved in random motions over a
wide range of scales (e.g., Spangler, 1999). It is still impossible to cover the
whole range of parameter variation (i.e., all the phases) in a single model of the ISM. In
this paper I shall review models of the diffuse phases of the ISM with densities below
10--$100\cmcube$. Simulations of MHD turbulence in dense clouds have been
discussed by Padoan et al.\ (1997, 1998), Stone et al.\ (1998), Ostriker et al.\ (1999)
and Heitsch et al.\ (1999); an extensive review and references can be found in
V\'azquez-Semadeni et al.\ (2000).

\section{Models of the multiphase ISM driven by supernovae}
In this paper we discuss the results of three models of the ISM
briefly summarized in Table~\ref{Models}. The models include
similar physical effects and their results agree in many respects. All are
non-linear, hydrodynamic or magnetohydrodynamic models that include
external gravity (typical of the Solar neighbourhood), suitably
idealized heating sources and radiative cooling (assuming partially
ionized, optically thin gas). All models employ artificial
viscosities to avoid excessively large gradients in the simulated
quantities. Neither of the models includes explicitly ionization
balance and cosmic rays.

\begin{table}[h]
\begin{center}
\caption{Recent comprehensive numerical models of the ISM driven by stellar winds and
supernovae.}   \label{Models}
\begin{tabular}{lccc}
\hline
&RBN$^{\rm(a)}$ &VSPP$^{\rm(b)}$ &KBSTN$^{\rm(c)}$ \\
\hline
Dimension       &2D: $xy$ and $xz$              &2D: $xy$      &3D: $xyz$\\
Box size, kpc   &$2\times2$ and $2\times15$     &$1\times1$
                                                &$0.5\times0.5\times2$\\
Resolution, pc  &10 and $10\times(10$--1000)    &2              &8\\
Heating    &SN~II, winds    &Diffuse, winds &SN~II, SN~I\\
Temperature range, K    &$10^2$--$10^8$ &$10^2$--$4\times10^4$
                                                &$10^2$--$10^8$\\
Magnetic fields &--             &imposed        &self-excited\\
Self-gravity    &--             &yes            &--\\
Rotation        &--             &yes            &yes\\
Cosmic Rays     &--             &--             &--\\
Ionization Balance &--          &--             &--\\
\hline
\end{tabular}
\end{center}
\begin{tabular}{l}
{\bf References}\\
$^{\rm(a)}$Rosen et al.\ (1993, 1996), Rosen \& Bregman (1995);\\
$^{\rm(b)}$V\'azquez-Semadeni et al.\ (1995), Passot et al.\ (1995),
Ballesteros-Paredes et al.\\
\hspace{1em} (1999);\\
$^{\rm(c)}$Korpi et al.\ (1999a,b).
\end{tabular}
\end{table}

The model of Rosen et al.\ (1993, 1996) and Rosen \& Bregman (1995)
(labelled RBN) uses a finite-difference ZEUS code in
two dimensions and neglects magnetic fields and rotation. These
authors have results for a region in the Galactic plane ($xy$) and in
a vertical plane ($xz$); the mesh size in the $z$-coordinate is
variable, growing with $z$ from $10\ppc$ to $1\kpc$. A
unique feature of this model is that the stellar Population~I is included
explicitly via equations for the stellar fluid coupled to the gas.  The heating sources
are Type~II SNe, which inject energy instantaneously, and
stellar winds modelled as a continuous heat source; both are correlated with the stellar
density. To alleviate numerical problems, the simulations have been split into
alternating stages, the interaction phase of the gas and stars when any motions are
neglected, and a stage of hydrodynamic evolution when any coupling between stars and gas
is neglected. SN explosions produce very hot gas, so the model contains three phases,
i.e., the hot, warm and cool gas.

V\'azquez-Semadeni et al.\ (1995, 1996, 1997, 1998), Passot et al.\
(1995), Scalo et al.\ (1998) and Ballesteros-Paredes et al.\ (1999),
labelled VSPP in Table~\ref{Models} (see also earlier results in
Passot et al., 1988, L\'eorat et al., 1990, and V\'azquez-Semadeni \&
Gazol, 1995), employ a pseudo-spectral numerical scheme and
restrict themselves to two dimensions in the Galactic plane. Their
best runs have a spatial resolution of $2\ppc$ or less, but only
features at scales above $\simeq10\ppc$ are reliable (Passot et al.,
1995).  To avoid numerical difficulties, this model includes diffuse heating (which
is thought to model the UV heating of the ISM) and stellar heating, but each stellar
energy release event is spread over a period of $6\times10^6\yr$. Furthermore, both the
heating and cooling rates are reduced by a factor 7--10 below the realistic values. As a
result, only the warm and cool gas phases are generated, with the maximum gas temperature
of $4\times10^4\K$. A peculiar feature of this model is the inclusion of self-gravity;
however, its r\^ole is only marginal at the densities reached in the simulations. The
model includes magnetic field, but any dynamo action is precluded in two dimensions.

The model of Korpi et al.\ (1999a,b) (KBSTN) is fully
three-dimensional and includes the effects of rotation and magnetic
field at a resolution of $8\ppc$.  This is the only of the three models
admitting realistic behaviour of vorticity and magnetic field including vortex
stretching and dynamo action.
The gas is heated by SNe modelled as thermal energy release over one time step which
can be as short as 10--100\yr. As a result, the gas is heated to $T\simeq10^8K$ at the
explosion site and hot, warm and cool gas phases can be identified in the simulations.
Both Type I and II SNe are included, which differ in both the occurrence rate and the
vertical distribution. The SN explosions occur at randomly chosen sites, but Type~I SNe
can only explode in those regions where the gas density exceeds the average at that
height. With this prescription, about 70\% of the Type~I SNe are clustered resembling OB
associations. The simulations have only been run over time span of the order of
$10^8\yr$, so it is yet unclear whether or not any magnetic fields can be supported by
dynamo action in this model. A related model is presented by Gudiksen (1999) where an
intricate algorithm for choosing SN explosion sites involving modelling of the stellar
initial mass function has been developed.

Despite different numerical approaches employed and  certain
differences in the physical content, these models yield many coherent
results which appear to be model-independent and which are our
subject in what follows.

\section{The multi-phase structure and cool gas
filaments}\label{Phases}
In all the models, the cooling function is truncated at
$T=100$--300\,K to avoid thermal instability at $T<10^5\K$.
Self-gravity is either neglected or unimportant.  Nevertheless,
cool, dense gas clouds with $T\la10^3\K$, $n=1$--$100\cmcube$ are a typical
feature of all the simulations. The clouds are elongated and are
better described as filaments or sheets (which can be distinguished from filaments only
in 3D models); rounder structures occur at the intersections of the filaments. The
filamentary $\HI$ clouds occur at positions of converging turbulent flow,
$\nabla\cdot\vect{v}<0$ and are not gravitationally bound (V\'azquez-Semadeni et al.,
1995; Ballesteros-Paredes et al., 1999). This indicates that they are formed by
compression (ram pressure) in the turbulent velocity field. The lifetime of the filaments
is shorter than $10^8\yr$ (Rosen \& Bregman, 1995), which is close to the kinematic time
scale. The filaments are mainly destroyed by mutual collisions. In models with SNe, a
more obvious mechanism of the cloud formation is compression in expanding shocks (Korpi
et al., 1999). The length of the filaments is 50--100\,pc and they are often extended
vertically because of the vertical outflow of the hot gas, resembling $\HI$ `worms'
observed in spiral galaxies (Rosen \& Bregman, 1995; Korpi et al., 1999). The
three-dimensional model of Korpi et al.\ (1999) has both filaments and sheets of cool
gas. In the model of Rosen \& Bregman (1995), where SNe are confined to a rather thin
layer, the cool filaments are not uncommon at $z=1$--3\,kpc, but in the model of Korpi et
al.\ (1999) they are confined to $|z|\la100\ppc$ because Type~II SNe that can occur at
large heights prevent cool gas from rising higher.

As could be expected, the hot phase ($T>10^5\K$) can only be
generated by the SNe. The hot gas fills bubbles surrounded by dense
shells; it breaks through the warm layer and streams into the halo
(or beyond the upper boundary of the computational domain) at
systematic velocities of 100--$200\kms$ (Korpi et al., 1999); Rosen \&
Bregman (1995) obtain $400\kms$ at $z=2$--3\,kpc.

The volume filling factors
of the hot, warm and cold phases are 40--30\%, 40--60\% and 20--10\%
at $z=1$--3\,kpc in the model of Rosen \& Bregman (1995). The filling factor of the
hot gas increases at larger heights at the expense of the other phases. The filling
factor of the hot phase in the model of Korpi et al.\ (1999) grows from 20--30\% at $z=0$
to 80--100\% at $|z|=1\kpc$; the remaining part of the volume is mostly occupied by the
warm gas. Unlike the hot gas, the warm component is, on average, in hydrostatic
equilibrium with a scale height of 200\,pc; this agrees with a detailed analysis
of hydrostatic equilibrium in the Solar vicinity by Fletcher \& Shukurov (1999). As
mentioned above, the cold gas is confined to $|z|\la100\ppc$. Rosen \& Bregman (1995)
perform detailed fitting of the vertical gas distribution in different phases. The scale
heights of the cold, warm and hot gas in one of their models (Run~E) are 225, 550 and
2000\,pc, respectively. Note that the gas called `cold' in the model, can be identified
with the cold and warm neutral medium, whereas the `warm' gas corresponds to the warm
ionized medium.

The phases of the modelled ISM are in a rough thermal pressure equilibrium. For example,
density and temperature span four orders of magnitude in the model of Korpi et al.\
(1999), but thermal pressure varies in space only by an order of magnitude. Similarly,
the density contrast is a factor of 50 whereas the pressure contrast is only a factor of 5
in the simulations of V\'azquez-Semadeni et al.\ (1995). The approximate pressure
equilibrium is not trivial to explain since the sound crossing time is not shorter than
the dynamic time scale because both the turbulent and systematic motions are transonic or
supersonic. V\'azquez-Semadeni et al.\ (1995) discuss how the pressure balance can result
from the instantaneous establishment of thermal balance in the gas. With their diffuse
heating, both the cooling time and the heating time are 10--100 times shorter than the
sound crossing time and the dynamic time scale, $10\ppc/10\kms\simeq10^7\yr$. Therefore,
the heating rate $\Gamma$ must be always close to the cooling rate $\rho\Lambda$ (with
$\rho$ the gas density); for $\Lambda\propto T^m$ (with $m$ a temperature-dependent
index), the equilibrium temperature follows as $T_{\rm eq}\propto(\Gamma/\rho)^{1/m}$.
The resulting pressure (assuming ideal gas) is $P_{\rm
eq}\propto\Gamma^{1/m}\rho^\gamma$, where $\gamma=1-1/m=0.33$--0.66 for $m=1.5$--2.9.
Since $\gamma<1$, denser regions are cooler, and this results in weaker spatial
variations of pressure (V\'azquez-Semadeni, 1998;
V\'azquez-Semadeni et al., 2000).  This explanation relies on a
non-localized nature of the heating (it also applies if the heating rate
is weakly dependent on density---Passot et al., 1995), but it may be
of more general importance.

\section{Global variability} \label{Var}
Quasi-periodic variations in the total thermal and kinetic energy (and also magnetic and
gravitational energy where appropriate) accompanied by similar variations in star
formation intensity are typical of the models. The period of the variations changes from
one model to another in the range (0.4--$4)\times10^8\yr$. Rosen \& Bregman, 1995, do not
discuss the global variability but mention oscillations, at a period $2.5\times10^8\yr$,
of the extent of the region occupied by filamentary gas structures. This is compatible
with the vertical sound crossing time and free-fall time across 1\,kpc, so these may be a
global acoustic mode, although V\'azquez-Semadeni et al.\ (1995) interpret the
variations as an analogue of spiral density waves in their model.

\section{Turbulence in the multiphase ISM}
The models include a random element in the heating source, so it is
not surprising that the gas motion is random, resembling developed
turbulence. V\'azquez-Semadeni et al.\ (1995, 1997) show that
compressible motions with a kinetic energy spectrum $E_k\propto
k^{-2}$ characteristic of shock waves contribute significantly to the
motions. Explosive heating by SNe driving shock waves would further
increase this contribution. These authors argue that the velocity
dispersion--size relation for interstellar clouds, $\Delta v\propto
R^{1/2}$ may result not from any kind of virial equilibrium, but
rather directly from the turbulent spectrum: $v_k^2\simeq kE_k\propto
k^{-1}$, where $v_k$ is velocity at a scale $R\propto k^{-1}$.

The difference between cloud formation mechanisms in
expanding SN shells and in colliding turbulent streams becomes
significant when the vorticity of the motions, $\omega$, is
considered.  Cool clouds formed by obliquely colliding streams of the
turbulent flow must have enhanced vorticity: V\'azquez-Semadeni et
al.\ (1995) obtain an enhancement by a factor of 10 for denser
clouds. Such an enhancement is not observed in a 3D model of Korpi et
al.\ (1999a,b) where gas filaments and sheets can be also formed by
expanding SN remnants and where essentially three-dimensional mechanisms are efficient in
amplifying vorticity in all the phases. Vorticity is amplified at curved shock fronts
distorted by ambient density inhomogeneities, by vortex stretching and by the baroclinic
effect, and 60--90\% of the turbulent energy is in vortical motions (Korpi et al.,
1999a). Vortex stretching is an essentially three-dimensional effect; furthermore, the
tendency for conservation of $\omega/\rho$ in a two-dimensional flow also contributes to
an artificial enhancement of vorticity in denser regions of a 2D model
(V\'azquez-Semadeni et al., 1995).

Korpi et al.\ (1999) computed the autocorrelation function of the vertical velocity for
the cold, warm and hot phases separately. The total (three-dimensional) r.m.s.\ turbulent
velocity of the cool gas is as low as $3\kms$, whereas it is close to $10\kms$ for the
warm gas (approximately the speed of sound in the warm phase). The correlation radius in
the warm gas is about 30\,pc (this is half the eddy size) and it only weakly changes with
$z$; this is interpreted as an indication of approximate vertical hydrostatic equilibrium
of the warm gas with a scale height of about 200\,pc. The r.m.s.\ random velocity in the
hot gas, $40\kms$ at $|z|\la1\kpc$, is significantly larger than in the other phases, but
still remains significantly smaller than the sound speed in the hot phase, $100\kms$. The
vertical streaming of the hot gas discussed above can feed turbulence higher in the halo,
so the turbulent velocity of $60\kms$ observed in the halo by Kalberla et al.\ (1998; see
also Pietz et al., 1998) seems to be perfectly compatible with the model. (Turbulent
velocities of order $100\kms$ in the halo have been invoked in galactic dynamo models
involving the halo---Sokoloff \& Shukurov, 1990; Brandenburg et al., 1992, 1993.) The
typical horizontal  radius of the hot regions at the midplane is about 20\,pc; this can
be identified with a chimney radius. The correlation radius for the hot phase grows with
$z$ together with the volume filling factor.

\section{Magnetic fields and dynamo action}
The only model admitting hydromagnetic dynamo action among the three
models discussed is that of Korpi et al.\ (1999a,b). These simulations
are initialized with a weak ($0.1\mkG$) regular magnetic field to provide a seed field
for the dynamo. Both the large-scale and the random magnetic fields show a
tendency to grow in the model, but runs spanning at least $10^9\yr$ are required to
demonstrate any dynamo action convincingly.

Magnetic fields in the 2D simulations of Passot et
al.\ (1995) do not decay only because the model has an imposed
uniform magnetic field, maintained throughout the simulations, which is tangled by
small-scale motions. Magnetic fields in the cool filaments formed by converging flows
often have a reversal along the filament axis. Magnetic energy in these simulations is
close to the kinetic energy in solenoidal motions at all scales. Magnetic fields, like
vorticity, are especially sensitive to the dimensionality of the model, and
two-dimensional results should be considered with caution.

According to dynamo theory, the growth time of the random magnetic
field due to the small-scale dynamo\footnote{In this context, `small'
scales are understood as those smaller than the correlation scale of
the turbulence, 50--100\,pc in the warm gas.} action is expected to be as
short as the eddy turnover time, $\simeq10^7\yr$ in the warm phase. A
plausible statistically steady state resulting from the saturation of
the dynamo action is an ensemble of magnetic flux ropes (e.g.,
Zeldovich et al., 1983, Sect.\ 8.IV). The length of the rope is of
the order of the turbulent scale, $l\simeq50$--$100\ppc$,
and the thickness is of the order of $l R_{\rm m}^{-1/2}$, where
$R_{\rm m}$ is the magnetic Reynolds number. Dynamo action can
occur provided $R_{\rm m}>R_{\rm m,cr}$, where the critical magnetic
Reynolds number is estimated as $R_{\rm m,cr}=60$--100 in
simplified models of homogeneous, incompressible turbulence.

Subramanian (1999) suggested that a steady state, reached via the back-action of the
magnetic field on the flow, can be established by the reduction of the
effective magnetic Reynolds number $R_{\rm m,eff}$ down to the value critical for the
dynamo action. Then  $R_{\rm m,eff}\approx R_{\rm m,cr}$ in the
steady state. Therefore, the thickness of the ropes in the steady
state can be estimated as $lR_{\rm m,cr}^{-1/2}$. Using a model
nonlinearity in the induction equation with incompressible velocity
field, Subramanian (1999) showed that the magnetic field strength
within the ropes $b_0$ saturates at the equipartition level with
kinetic energy density, $b_0^2/8\pi\simeq\half\rho v^2$, where $\rho$ is the gas density
and $v$ is the r.m.s.\ turbulent velocity. The average magnetic energy density is
estimated as $\langle b^2/8\pi\rangle\simeq CR_{\rm m,cr}^{-1}\half\rho v^2$, where $C$
is a numerical coefficient, presumably of order unity. The volume filling factor of the
ropes then results as $f\simeq R^{-1}_{\rm m,cr}\simeq0.01$; correspondingly, the mean
magnetic energy generated by the small-scale dynamo in the steady state is about 1\% of
the turbulent kinetic energy density.

Using parameters typical of the warm phase of the ISM, this theory
predicts that the small-scale dynamo would produce magnetic flux
ropes of the length (or the curvature radius) of about
$l\simeq50$--100\,pc and thickness $lR_{\rm
m,cr}^{-1/2}\simeq5$--$10\ppc$. The field strength within the ropes, if
at equipartition with the turbulent energy, has to be of order $1.5\mkG$ in the warm phase
($n=0.1\cmcube$, $v=10\kms$) and $0.5\mkG$ in the hot gas ($n=10^{-3}\cmcube$,
$v=40\kms$). Note that other models of the small-scale dynamo admit solutions with
magnetic field strength within the ropes being significantly above the equipartition
level (Belyanin et al., 1993).

The small-scale dynamo is not the only mechanism producing turbulent
magnetic fields (e.g., Beck et al., 1996, Sect.\ 4.1 and references therein). Any
mean-field dynamo action producing magnetic fields at scales exceeding the turbulent
scale also generates small-scale, turbulent magnetic fields. Similarly to the mean
magnetic field,  this component of the turbulent field presumably has a filling factor
close to unity in the warm gas and its strength is expected to be close to
equipartition with the turbulent energy at all scales.  (As argued below, this component
of the turbulent magnetic field may be confined to the warm gas, so magnetic field in
the hot phase may have a better pronounced ropy structure.)

The overall structure of the
interstellar turbulent magnetic field in the warm gas can be envisaged as a quasi-uniform
fluctuating background with one percent of the volume occupied by
flux ropes (filaments) of a length 50--100\,pc with well-ordered magnetic field.
The ropes can plausibly produce elongated, localized filamentary structures of
polarized emission and Faraday rotation recently observed in the Milky Way by Wieringa et
al.\ (1993) and Gray et al.\ (1998). The basic distribution described would be further
complicated by compressibility, shock waves, MHD instabilities (such as Parker
instability), etc.

The site of the mean-field dynamo action is plausibly the warm phase rather
than the other phases of the ISM. The warm gas has a large filling factor (so it can
occupy a simply connected global region), it is, on average, in a state of hydrostatic
equilibrium, so it is an ideal site for both the small-scale and mean-field dynamo
action. Molecular clouds and dense $\HI$ clouds have too small a filling factor to be of
global importance. Fletcher \& Shukurov (1999) argue that, globally, molecular clouds can
be only weakly coupled to the magnetic field in the diffuse gas. (However, the
small-scale dynamo can work {\it within\/} the clouds.) The time scale of the small-scale
dynamo in the hot phase is $\ga l/v\simeq10^6\yr$ for $l=40\ppc$ and $v=40\kms$. This can
be shorter than the advection time due to the vertical streaming, $h/V_z\simeq10^7\yr$
with $h=1\kpc$ and $V_z=100\kms$. Therefore, the small-scale dynamo action should be
possible in the hot gas even at small heights. However, the growth time of the mean
magnetic field must be significantly longer than $l/v$ reaching a few hundred Myr (e.g.,
Beck et al., 1996, Sect.\ 4.4). Thus, the hot gas can hardly contribute significantly to
the the mean-field dynamo action in the disc and can drive the dynamo only in the halo.
The fountain flow rather pumps the mean magnetic field out from the disc (Brandenburg et
al., 1995); this effect can contribute to the saturation of the mean-field dynamo in the
disc.

\section*{References}
\references\noindent

Ballesteros-Paredes J., V\'azquez-Semadeni E., Scalo J.
\Journal{1999}{\ApJ}{515}{286}.

Beck R., Brandenburg A., Moss D., Shukurov A., Sokoloff D.
\Journal{1996}{\ARAaAp}{34}{155}.

Belyanin M., Sokoloff D., Shukurov A. (1993) {\it Geophys.\
Astrophys.\ Fluid Dyn.}  {\bf 68}, 237.

Brandenburg A., Donner K.J., Moss D., Shukurov A., Sokoloff D.D., Tuominen I.
\Journal{1992}{\AAp}{259}{453}.

Brandenburg A., Donner K.J., Moss D., Shukurov A., Sokoloff D.D., Tuominen I.
\Journal{1993}{\AAp}{271}{36}.

Brandenburg A., Moss D., Shukurov A. \Journal{1995}{\MNRAS}{276}{651}.

Fletcher A., Shukurov A. (1999) this volume.

Gray A.D., Landecker T.L., Dewdney P.E., Taylor A.R. \Journal{1998}{\Nat}{393}{660}.

Gudiksen B.V., {\it The Interstellar Medium in Disc Galaxies; the Influence of SNe and
Magnetic Fields}, Master Thesis (Astron.\ Obs., Copenhagen Univ.\ 1999).

Heitsch F., Mac Low M.-M., Klessen R. (1999) this volume.

Kalberla P.M.W., Westphalen G., Mebold U., Hartmann D., Burton W.B.
\Journal{1998}{\AAp}{332}{L61}.

Korpi M.J., Brandenburg A., Shukurov A., Tuominen I. (1999a)
in {\it Interstellar Turbulence. Proc.\ 2nd Guillermo Haro Conf.,
Puebla, Mexico, Jan.\ 12--16, 1998}, eds.\ Franco J.\ \& Carrami\~nana A. (Cambridge
Univ.\ Press), p.~127.

Korpi M.J., Brandenburg A., Shukurov A., Tuominen
I., Nordlund \AA. \Journal{1999b}{\ApJL}{514}{L99}.

L\'eorat J., Passot T., Pouquet A. \Journal{1990}{\MNRAS}{243}{293}.

Ostriker E.C., Gammie C.F., Stone J.M. \Journal{1999}{\ApJ}{513}{259}.

Padoan P., Jones B.J.T., Nordlund \AA. \Journal{1997}{\ApJ}{474}{730}.

Padoan P., Juvela M., Bally J., Nordlund \AA. \Journal{1998}{\ApJ}{504}{300}.

Passot T., Pouquet A., Woodward P. \Journal{1988}{\AAp}{197}{228}.

Passot T., V\'azquez-Semadeni E., Pouquet A. \Journal{1995}{\ApJ}{455}{536}.

Pietz J., Kerp J., Kalberla P.M.W., Burton W.B., Hartmann D., Mebold U.
\Journal{1998}{\AAp}{332}{55}

Rosen A., Bregman J.N. \Journal{1995}{\ApJ}{440}{634}.

Rosen A., Bregman J.N., Norman M.L. \Journal{1993}{\ApJ}{413}{137}.

Rosen A., Bregman J.N., Kelson D.D. \Journal{1996}{\ApJ}{470}{839}.

Scalo J., V\'azquez-Semadeni E., Chapell D., Passot T.
\Journal{1998}{\ApJ}{504}{835}.

Sokoloff D., Shukurov A. \Journal{1990}{\Nat}{347}{51}.

Spangler S. (1999) this volume.

Stone J.M., Ostriker E.C., Gammie C.F. \Journal{1998}{\ApJL}{508}{L99}.

Subramanian K. (1999) {\it Phys.\ Rev.\ Lett.}, in press.


V\'azquez-Semadeni E., Gazol A. \Journal{1995}{\AAp}{303}{204}.

V\'azquez-Semadeni E., Passot T., Pouquet A. \Journal{1995}{\ApJ}{441}{702}.

V\'azquez-Semadeni E., Passot T., Pouquet A. \Journal{1996}{\ApJ}{473}{881}.

V\'azquez-Semadeni E., Ballesteros-Paredes J.,
Rodr\'{\i}quez L.F., \Journal{1997}{\ApJ}{474}{292}.

V\'azquez-Semadeni E., Cant\'o J., Lizano S. \Journal{1998}{\AAp}{492}{596}.

V\'azquez-Semadeni E., Ostriker E., Passot T., Gammie C.F., Stone J.M. (2000) in {\it
Protostars and Planets IV,\/} eds.\ Mannings V., Boss A.P. \& Russell
S.S. (Univ.\ Arizona Press, Tucson), in press.

Wieringa M.H., de Bruyn A.G., Jansen D., Brouw W.N., Katgert P.
\Journal{1993}{\AAp}{268}{215}.

Zeldovich Ya.B., Ruzmaikin A.A., Sokoloff D.D., {\it Magnetic Fields in
Astrophysics\/} (Gordon and Breach, New York 1983).

\end{document}